
\magnification\magstep1
\hsize5.45truein
\hoffset.8truein
\baselineskip24truept
\topskip10truept
\def\SE{\langle T_{\mu\nu}\rangle}
\def\L{\hbox{{\it \$}}}
{
\nopagenumbers
\hfill IP-ASTP-12-93

\hfill May, 1993

\null
\vfill

\centerline{\bf The Energy Density of a Gas of Photons}
\centerline{\bf Surrounding a Spherical Mass $M$ at a Non-Zero Temperature}
\vskip.5in

\centerline{\it Achilles D. Speliotopoulos}
\footnote{}{\rm Bitnet address: PHADS@TWNAS886}

\vskip.5cm

\centerline{\it Institute of Physics}
\centerline{\it Academia Sinica}
\centerline{\it Nankang, Taipei, Taiwan 11529}

\vfill

\centerline{\bf Abstract}

\vskip.25in

{
 \baselineskip26truept
 \leftskip2in
 \rightskip2in
 \noindent The equations determining the energy density $\rho$ of a gas of
 photons in thermodynamic equilibrium with a spherical mass $M$ at a
 non-zero temperature  $T_s>0$ is derived from Einstein's equations.
 It is found that for large $r$, $\rho \sim 1/r^2$ where the
 proportionality constant is a {\it fundamental constant\/} and is
 the same for all spherical masses at all temperatures.
}
\vskip1truecm
\vfill
\supereject
}
\pageno=2
\noindent{\bf \S 1. Introduction}

In the standard Schwarzchild solution $[1]$ of Einstein's equations for a
static, spherical geometry the spacetime outside of a sphere of mass
$M$ and radius $R$ is taken to be a vacuum; empty and free of
particles. This is certainly true if the sphere is at absolute zero
temperature. When, however, the sphere has a non-zero temperature,
then we would expect, on a physically basis, a gas of thermal photons to be
present. If the system is in thermodynamic equilibrium, then the
spacetime surrounding the sphere will not be empty, but
will instead be filled with blackbody radiation. As these thermal
photons also have a certain non-zero energy, we would expect the
presence of this blackbody radiation to also contribute towards
determining the geometry of the spacetime.

For any physically reasonable temperatures the energy density of the
photons is very small in comparison to the mass density of the sphere
and the presence of the photons is ignored in the usual Schwarzchild
analysis. This approximation is certainly valid near the spherical
body but what happens when one is very far away from the sphere? One
should remember that the Schwarzchild solution is asymptotically
flat. Due to the mass $M$ of the body being confined within its radius
$R$, as one goes further and further away from the sphere the affect
of its mass $M$ on the curvature of spacetime becomes less and less.
The blackbody radiation, on the other hand, is unconfined. It extends
from the surface of the sphere to fill the rest of the spacetime.
Although we would expect the energy density of the photons to also
decrease as one moves further and further away from the sphere, the
important question is how fast it will do so. Let us, for the moment,
consider a sphere of radius $r>R$ and the total amount of energy
within it. When $r$ is near $R$, we would expect that the fraction
$\epsilon_M$ of this energy which is due to the mass $M$ will be much
greater than $\epsilon_\gamma$, the fraction of the total energy
which is due to the photons. Consequently, we would expect the mass
$M$ to be the dominant factor in determining the geometry of the
spacetime within $r$ and would expect the
Schwarzchild solutions to be valid in this region. As, however, $r$
increases, $\epsilon_M$ decreases since the mass $M$ of the body is
fixed, while $\epsilon_\gamma$ increases since the blackbody
radiation extends throughout the spacetime. If the energy density of the
photons decreases sufficiently rapidly so that
$\epsilon_\gamma\ll\epsilon_M$ for all $r$, then the mass $M$ will
always be the dominant factor in determining the geometry of  the
spacetime. Although the Schwarzchild solution would be modified
somewhat by the presence of the photons, we would not expect these
modifications to be very drastic. In particular, the spacetime should
still be asymptotically flat. If, on the other hand, the energy density
does not decrease rapidly enough and $\epsilon_\gamma \gg \epsilon_M$
for $r$ larger than some $r_0$, then we would expect that in this
region of spacetime it is the photons which will determine the
spacetime geometry. In particular, we would expect the solutions to
Einstein's equations in this regime to be very much different from
the Schwarzchild solution. The spacetime may not even be asymptotically
flat.

In this paper we shall study some of the affects of non-zero
temperatures on the static, spherically symmetric solutions of
Einstein's equations. The system we shall be considering consists of
a spherical body with a mass $M$ and a radius $R$ which, due to the
sphere being at a temperature $T_s>0$, is surrounded by blackbody
radiation. The system as a whole will be assumed to be in
thermodynamic equilibrium with the body serving as the heat reservior
for the system. In particular, this means that the spherical body
is assumed to be in thermodynamic equilibrium with the photons
surrounding it. It is moreover assumed that the body has not
undergone complete gravitational collapse into a blackhole, is
non-rotating, and is not electrically charged. Nor shall there be any
other massive objects present in this spacetime. The only difference
between this system and the one studied by Schwarzchild is the
presence of a non-zero temperature for the sphere.

Our aim in this paper is two fold. First, we shall
derive a set of coupled differential equations which will determine the
total energy density of the photons. As these are non-linear
equations, we shall not be able to solve them analytically. We shall,
nonetheless, be able to obtain both asymptotic $r\to\infty$
as well as $r\to R$ solutions to them. Note, however, that we shall
be determining the photon's {\it total\/} energy density in the
gravitational field, and {\it not\/} its blackbody spectrum. Second,
we shall show that the geometry of this spacetime differs drastically
from the Schwarzchild geometry at large $r$. In fact with the
presence of the photons the spacetime is no longer asymptotically
flat.

In order to avoid working with non-equilibrium systems, we have
assumed that the mass $M$ is in thermodynamic equilibrium
with the blackbody radiation surrounding it. Unfortunately, even
equilibrium quantum statistical mechanics on curved spacetimes has
yet to be satisfactorily formulated. The closest that we have come to
a complete formalism is that given in $[2]$. For various reasons,
however, it will be difficult, if not impossible, to analyze the
system in the manner outlined therein and it is fortunate that all
that we shall need is the Tolman-Oppenheimer-Volkoff equation for
hydrostatic equilibrium. This, combined with the observation that the
energy-momentum tensor of a gas of pure photons is traceless, shall be
sufficient to determine the energy density almost uniquely. This
method of deriving the energy density has the added advantage of
not only taking into account the affects of the curvature of
spacetime on the photons, but also the reciprocal affect of the
photons on the spacetime curvature. Photons are {\it
not\/} treated as test particles in our analysis. In this, and other,
ways our method differs from that discribed in $[2]$. The only
difficulty that we shall encounter is when we try to identify the
temperature of the system. Since we do not have an established
formalism which will automatically do this for us, we shall, in the
end, have to rely on other physical arguments to do so.

In the formalism given in $[2]$, as indeed in most treatments
of thermodynamics in general relativity, the temperature of the system
is taken to be a constant throughout the spacetime. This is, it would
seem to us, an oversimplification, for the following reason. For
massless particles the temperature of the system at equilibrium may
be interpreted physically as the most probable energy that any one
particle in the statistical ensemble may have. (The case of massive
particles is much more complicated and will not be considered here.)
In a gravitational field this should include not only its kinetic
energy, but also its gravitational energy as well. The two cannot be
seperated covariantly. Indeed, it is known that the frequency of a
photon, and thus its energy, when measured at different points in a
gravitational field will either be ``redshifted'' or ``blueshifted''
with respect to one another. We would on this basis expect that
temperature too should vary from point to point on the manifold.

\noindent{\bf \S 2. The Hydrostatic Equation}

We begin with an $N$ dimensional manifold ${\bf M}$ with a
metric $g_{\mu\nu}$ which has a signiture of $(-,+,+,+)$. Greek
indices shall run from $0$ to $N-1$ and the summation convention is
used throughout. It is further assumed that $g_{\mu\nu}$ is static,
meaning that there exists a timelike Killing vector $\xi_\mu$ for the
system. We shall also assume that the system contains only one heat
reservior.

Next, let $T_{\mu\nu}$ be an energy momentum tensor operator defined on
$\bf M$. We shall, in a semi-classical approximation, treat
$g_{\mu\nu}$ as a background, {\it classical\/} field.  We next
denote the thermodynamic average of $T_{\mu\nu}$ by $\langle
T_{\mu\nu}\rangle$. We shall not need a specific definition of this
average, but rather that it satisfy a few basic properties that we
would expect from {\it any\/} equilibrium thermodynamic average.
First, it should be ``time independent'', meaning that
$$
\L_\xi \SE = 0,
\eqno(1)
$$
where $\L_\xi$ denotes the Lie derivative along the $\xi_\mu$
direction. It is for this reason that we required $\bf M$ to have a
timelike Killing vector. Physically, it means that the background
field $g_{\mu\nu}$ cannot change with respect to time the total
energy contained in the matter fields so that the system as a whole
can be in equilibrium. Second, we require that the average be
anomaly-free
$$
\nabla_\lambda \langle T_{\mu\nu}\rangle = \langle \nabla_\lambda
T_{\mu\nu}\rangle,
\eqno(2)
$$
where $\nabla_\lambda$ denotes the covariant derivative.
Third, after the average is taken the energy momentum
tensor should have the form $\langle T_{\mu\nu}\rangle = \rho u_\mu
u_\nu + p (g_{\mu\nu} + u_\mu u_\nu)$ were $\rho$ and $p$ are the
proper energy density and pressure, respectively, and $u_\mu$ is a unit
velocity vector which must lie in the $\xi_\mu$ direction if $(1)$ is
to hold. We shall make this dependency explicit by writting
$$
\SE = -\rho {\xi_\mu\xi_\nu\over \xi^2} + p \left(g_{\mu\nu} -
{\xi_\mu\xi_\nu\over \xi^2}\right).
\eqno(3)
$$
Then $(1)$ requires that
$$
\xi^\lambda \nabla_\lambda \rho = 0 \qquad,\qquad \xi^\lambda
\nabla_\lambda p = 0,
\eqno(4)
$$
meaning that $\rho$ and $p$ are functions of vectors lying in the
$N-1$ dimensional hypersurface perpendicular to $\xi_\mu$ {\it and/or\/}
$\xi^2= \xi_\mu\xi^\mu$. Finally, since the system is conserved,
$\nabla^\mu\SE = 0$ and
$$
\nabla_\mu p + (\rho+p) {\nabla_\mu\vert\xi\vert\over\vert\xi\vert} =
0,
\eqno(5)
$$
where $\vert\xi\vert \equiv \sqrt{-\xi^2}$ and we have used $(3)$ and
Killing's equation
$$
\nabla_\mu \xi_\nu + \nabla_\nu\xi_\mu = 0,
\eqno(6)
$$
in obtaining $(5)$. This is the generalization of the
Tolman-Oppenheimer-Volkoff $[3]$ equation for hydrostatic equilibrium to
general, static spacetimes and it reduces to the usual hydrostatic
equation in the case of spherically symmetric spacetimes.

We next consider a region of spacetime which contains only photons.
In this region
$$
4\pi T_{\mu\nu} = F_{\mu\lambda} F^\lambda_\nu -
{1\over 4}g_{\mu\nu} F_{\alpha\beta}F^{\alpha\beta},
\eqno(7)
$$
where $F_{\mu\nu}$ is the field strength tensor. This energy-momentum
tensor operator is traceless $T_\mu^\mu = 0$ in four spacetime dimensions.
Consequently, $\langle T_\mu^\mu\rangle = 0$ and for photons $\rho =
(N-1)p$. The hydrostatic equation $(5)$ is now trivial to solve yielding
$$
\rho = {\sigma\over \vert\xi\vert^N},
\eqno(8)
$$
where $\sigma$ is an arbitrary {\it constant\/}. The average energy
momentum tensor for photons is thus given by
$$
\SE = {1\over (N-1)}{\sigma\over\vert\xi\vert^N}
        \left(g_{\mu\nu}-N{\xi_\mu\xi_\nu\over\xi^2}\right).
\eqno(9)
$$
To determine $\SE$ completely one must first determine $\xi_\mu$
for the manifold. As the presence of the photons will also affect the
curvature of the spacetime, determining $\xi_\mu$ ultimately involves
solving Einstein's equations using $(9)$ as the source term
$$
{1\over (N-1)}
{\sigma\over\vert\xi\vert^N}
        \left(
                g_{\mu\nu}-N{\xi_\mu\xi_\nu\over\xi^2}
        \right) = {1\over 8\pi} R_{\mu\nu},
\eqno(10)
$$
where $R_{\mu\nu}$ is the Ricci tensor and we are using geometrized
units in which $G=c=k_B=\hbar=1$. The $R_\mu^\mu$ term is absent
since $\SE$ is traceless.

The task now is to interpret $(8)$ physically. Let us
consider, for the moment, the case of Minkowski spacetime and enclose the
system in a very, very large box which is connected to a heat
reservior at a fixed temperature. Killing's
equation is now a simple partial differential equation and we may choose
a coordinate system in which its solution for a timelike Killing
vector is $\xi^f_\mu = (-\beta^f,0, 0, 0)$ where
$\beta^f$ is a constant and the superscript $f$ reminds us that this
is the Minkowski spacetime. From $(8)$ we find that in four
dimensions,
$$
\rho^f = {\sigma\over (\beta^f)^4},
\eqno(11)
$$
which, if we interpret $1/\beta^f$ as the temperature of the heat
reservior, is just Boltzmann's law for photons. Then $\sigma=
\pi^2k_B^4/(15\hbar^3c^3)$ is identified with the blackbody radiation
constant.

Using the Minkowski spacetime case as motivation, we shall tentatively identify
the temperature $T$ of the system
as
$$
T ={1\over \vert\xi\vert},
\eqno(12)
$$
in general, static spacetimes and see whether or not this will make
physical sense. First, we note that although $T$ does very with
position, it is ``time independent'', namely
$$
\L_\xi T = {T^3\over 2}\xi^\mu\nabla_\mu \xi^2 = 0,
\eqno(13)
$$
as one would expect for a system in equilibrium. Second, we note that
variations in $T$ are due solely to the gravitational field. In
fact, the temperature at various points of the manifold is related to
one another by just the redshift factor,
$$
{T(x)\over T(x')} =
\left(-\xi^2(x')\over-\xi^2(x)\right)^{1/2},
\eqno(14)
$$
which is precisely what one would expect from time dilation and the
frequency shift of photons in a gravitational field. Finally, we note
that timelike Killing vectors $\xi$ in non-rotating systems are determined
by the geometry of the spacetime only up to an overall constant, as
can be seen explicitly in Killing's equation $(6)$. Consequently,
we have the freedom to attach an overall constant to the Killing
vector which we can then identify as the inverse temperature of the heat
reservior. Indeed, from $(14)$ we see that {\it relative\/}
temperatures between two points on the manifold are determined solely
by geometry and to determine an {\it absolute\/} temperature requires
choosing a reference point on the manifold from which we can measure
all subsequent temperatures with respect to. The most natural
reference point to choose is the heat reservior of the system and to
measure all other temperatures based on its value.

The procedure for determining the energy density of thermal photons
in any static geometry is now clear. First we solve Killing's
equations up to an overall constant for a timelike Killing vector
$\xi_\mu$ in terms of the componants $g_{\mu\nu}$ of the metric of
the spacetime. To determine the overall
constant, we use as a boundary condition for $\xi_\mu$ the
temperature $T_{hr}$ of the heat reservior of the system by
evaluating $(12)$ at a point $x_{hr}$ on the surface of the heat
reservior. It is required that the heat reservior have a constant
temperature throughout its surface. Finally, Einstein's equations
$(10)$ are solved with the appropriate boundary conditions to
determine $g_{\mu\nu}$. These boundary conditions are usually also
given at the heat reservior. The energy density  $\rho$ of the
photons and the geometry of the spacetime are thus determined.

\noindent{\bf\S 3 Spherical Geometry}

We shall now attempt to solve $(10)$ for a static, spherical
geometry. Specifically, we shall
consider a non-rotating spherical body of mass $M$, and radius $R$
which at its surface has a temperature $T_s$. The spherical body will
serve as the heat reservior for the system. The most general metric
which is static and spherically symmetric has the form $[3]${}
$$
ds^2 = -f dt^2 + hdr^2 + r^2 d\theta^2 + r^2\sin^2\theta d\phi^2,
\eqno(15)
$$
where $f$ and $h$ are functions of $r$ only. Killing's equation is
now straightforward to solve giving $\xi_\mu = (-c_k f(r), 0, 0, 0)$,
and $\xi^2 = - c_k^2 f(r)$ for a timelike Killing vector. $c_k$ is an
arbitrary constant which is determined by evaluating $(12)$ at the
surface of the reservior: $1/c_k = T_sf(R)^{1/2}$.

As for the boundary conditions for $f$ and $h$, we choose
them in the following manner. Take a point $r>R$ and
consider the amount of energy contained within a sphere of this
radius which is due to the photons. Clearly, because of the
presence of the photons, the geometry of the spacetime within $r$
will be different from the Schwarzchild geometry. Let us, however,
take $r\to R^{+}$ so that the amount of energy contained in the
sphere due to the photons gradually decreases. Their affect on the
geometry of spacetime must also do so correspondingly and just
outside the body we would expect the geometry of the manifold to be
the same as that of the Schwarzchild geometry. Consequently, we
choose as our boundary conditions
$$
\lim_{r\to R^{+}} f(r) = 1-{2M\over R}\qquad,\qquad
\lim_{r\to R^{+}} {1\over h(r)} = 1-{2M\over R}.
\eqno(16)
$$

Einstein's equations given by $(10)$ are
$$
\eqalignno{
{8\pi\rho_0\over f^2}
        &
        =
        {1\over fh}
        \left[
                {f''\over 2}
                -
                {f'\over 4}
                        \left(
                                {f'\over f}+{h'\over h}
                        \right)
                +
                {f'\over r}
        \right],
        \cr
{8\pi\rho_0\over 3f^2}
        &
        =
        {1\over fh}
        \left[
                {-f''\over 2}
                +
                {f'\over 4}
                \left(
                        {f'\over f}+{h'\over h}
                \right) +
                {fh'\over r}
        \right],
        \cr
{8\pi\rho_0\over 3f^2}
        &
        =
        - {1\over 2rh}
        \left(
                {f'\over f}+{h'\over h}
        \right) +{h'\over r h^2} +
        {1\over r^2}\left(1-{1\over h}\right),
        &(17)
        \cr
}
$$
where $\rho_0 = \sigma T^4_s f^2(R)$ and the primes denote
derivatives with respect to $r$. They may be reduced to two coupled,
nonlinear differential equations
$$
\eqalignno{
{32\pi\rho_0\over 3f^2}
        =
        &
        {1\over rh}\left({f'\over f}+{h'\over h}\right)\>
        \cr
{8\pi\rho_0\over f^2}
        =
        &
        {h'\over r h^2} + {1\over r^2}\left(1-{1\over h}\right).
        &(18)
        \cr
}
$$
It is doubtful that these equations can be solved analytically.
There is, nonetheless, one general feature that we can determine
from these equations alone. Since we require
$h>0$ for all $r>R$, from $(18)$ we find that $f$ is a
monotonically increasing function of $r$. Consequently, $\rho =
\rho_0/f^2$ is a monotonically decreasing function of $r$. The energy
density of the photon gas is at its largest at the surface of the
sphere and decreases monotonically as one goes further and further
away from it. This is exactly what we would have expected physically.
Notice also that since $T=1/\vert\xi\vert= T_s[f(R)/f(r)]^{1/2}$, the
temperature of the photon gas also decreases monotonically with $r$.

We shall now obtain approximate solutions to $(18)$ in the small and
large $r$ limits. We first consider the near field solutions when $r$
is near $R$ and write $f\approx 1 - \Gamma$, and $h^{-1} \approx
1-\Lambda$ for $\Gamma$, $\Lambda \ll 1$. For the boundary conditions
to be consistant with this approximation, we shall require
$2M/R\ll1$. Then ignoring terms quadratic in $\Gamma$ and $\Lambda$,
we find that
$$
\eqalignno{
{1\over h(r)}
        &
        \approx 1 - {2M\over r}\left\{1 + {4\pi\rho_0 r^3 \over 3M} \left[1 -
        \left(R\over r\right)^4\right]\right\},
        \cr
f(r) &
        \approx
        1 - {2M\over r}\left\{1 - {4\pi\rho_0 r^3 \over 3M} \left[1 -
        \left(R\over r\right)^2\right]\right\},
        \cr
\rho(r) &
        =
        \rho_0
        \left\{
                1 - {2M\over r} +{8\pi\over 3} \rho_0 r^2 \left[1 -
                \left(R\over r\right)^2\right]
        \right\}^{-2}.
        &(19)
        \cr
}
$$
Notice, however, that $\Gamma$ and $\Lambda$ increases quadratically
with $r$ and at some point $r_m$ they will no longer be valid. To
estimate $r_m$, we enforce the condition that $\Gamma(r)\ll 1$ for
$r<r_m$. This gives
$$
{4\pi\rho_0\over 3} r_m^3 \approx M,
\eqno(20)
$$
as a determining equation for $r_m$. The near field solutions $(19)$ are
valid as long as the total energy of the photons confined in
a sphere of radius $r$ is much less than the mass of the spherical
body itself.

As for the asymptotic, $r\to \infty$ solutions, we obtain them in the
following manner. First we define
$$
\rho = {\Delta\over 4\pi r^2}\>, \qquad {1\over h} = 1-2K.
\eqno(21)
$$
Then $(18)$ may be written as
$$
\eqalignno{
{d\Delta\over dy} =
                &
                -{2\Delta\over 1-2K}\left({2\over 3} \Delta + 4K
                -1\right) \>
                \cr
{dK\over dy} =
                &
                \>\Delta - K \>,
                &(22)
                \cr
}
$$
where $y = \log(r/r_0)$ for some $r_0$. These differential equations
have a fix point at
$$
\Delta_a = K_a = {3\over 14}
\eqno(23)
$$
where the derivatives of $\Delta$ and $K$ vanish. Then perturbing about
this fix point,
$$
{d\>\>\over dy}\pmatrix{\Delta-\Delta_a\cr
                        K-K_a\cr}
                =
                \pmatrix{-1/2&-3\cr
                         1&-1\cr}
                \pmatrix{\Delta-\Delta_a\cr
                        K-K_a\cr}.
\eqno(24)
$$
This matrix has eigenvalues
$$
\lambda_\pm = -{3\over 4}\pm i{\sqrt {47}\over 4}
\eqno(25)
$$
so that the fix point $(23)$ is {\it stable}. Physically, this means that
no matter what initial conditions are chosen for $\Delta$ and $K$,
both functions will eventually flow to the fix point $(23)$ at large
enough $r$. We can see this explicitly by solving $(24)$ {}
$$
\eqalignno{
K(r) =
        &
        {3\over 14}
        \left\{1 + A
        \left(
                r_0\over r
        \right)^{3/4}
        \sin\left(
                {\sqrt{47}\over 4} \log{{r\over r_0}}
        \right),
        \right\}
        \cr
\Delta(r) =
        &
        {3\over 14}
        \Bigg\{
                1 + A\left(r_0\over r\right)^{3/4}
                \Bigg[
                        {\sqrt{47}\over 4}
                        \cos\left(
                                {\sqrt{47}\over 4}\log{{r\over r_0}}
                        \right)
        \cr
        &
                        +
                        {1\over4}
                        \sin\left(
                                {\sqrt{47}\over 4}\log{{r\over r_0}}
                        \right)
                        \Bigg]
        \Bigg\}.
       &(26)
        \cr
}
$$
{}$A$, and $r_0$ are constants which require
matching boundary conditions that are given at small $r$ to determine.
As the solutions to $(22)$ for intermediate $r$ are not known
analytically, we are not able to do this explicitly. Nevertheless,
numerical calculations, and a formal perturbative solution of $(18)$
treating $\rho_0$ as the perturbation indicates that $r_0\sim
1/\sqrt{\rho_0}$ as long as $1/\sqrt\rho_0 > r_m$. We would therefore
expect $(26)$ to hold whenever $r\gg 1/\sqrt{\rho_0}$. Note also
that solutions to $(22)$ approach the fix point $(23)$ very
slowly; basicly as $r^{-3/4}$.

In the very large $r$ limit solutions to $(22)$ asymptotically
approaches $f_a = (56\pi\rho_0/3)^{1/2} r$, and $h_a = 7/4$ where we
have used the subscript $a$ to denote the asymptotic solutions. Thus
the metric at large $r$ is
$$
ds^2 = -\left(56\pi\rho_0\over 3\right)^{1\over2}r\> dt^2 + {7\over
4}dr^2 + r^2 d\theta^2 + r^2\sin^2\theta d\phi^2.
\eqno(27)
$$
and we can now see explicitly that this spacetime is not
asymptotically flat. Next, we find that the asymptotic energy density
is
$$
\rho_a = {3c^4\over 56\pi G r^2},
\eqno(28)
$$
where we have replaced the correct factors of $c$ and $G$. At very
large $r$, the energy density decreases as $1/r^2$ with a
proportionality constant which is an universal number and is
independent of either the mass $M$ or the temperature $T_s$ of the
sphere. This is once again the consequence of $(22)$ being a
non-linear differential equation and having a stable, non-zero
fix point. The temperature of the photon gas in the
asymptotic limit is then
$$
k_B T(r) = \left(45 \hbar^3c^7\over 56\pi^3 G r^2\right)^{1/4}
         = m_{pl} c^2\left({45 \over 56\pi^3}{l_{pl}^2\over r^2}\right)^{1/4},
\eqno(29)
$$
where $m_{pl} = (\hbar c/ G)^{1/2}$ is the Planck mass and $l_{pl}=
m_{pl}G/c^2$ is the Planck length and we have explicitly used
$\sigma = \pi^2k_B^4/(15\hbar^3c^3)$. Although $m_{pl}$
is very large, one should remember that $(29)$ is valid only when $r$ is
also quite large. As $l_{pl}\sim 10^{-33}$cm, this ensures that
$k_BT(r)$ will always be very much smaller than the Planck energy. In
fact, $(18)$ gaurentees that for $r>R$,
$T(r)\le T_s$, the temperaure at the surface of the sphere. We should
also mention that the asymptotic solutions {\it are themselves\/}
solutions of $(22)$ at {\it any\/} $r$, as can be seen explicitly.
They do not, unfortunately, satisfy the correct boundary conditions.

We can also calculate the total average energy of the photons
$$
E = \int d^3x \sqrt{h} \rho = \int_R^\infty \sqrt{h} {dm\over dr} dr,
\eqno(30)
$$
from $(18)$. Since $m\sim r$ for large $r$, we do not expect this $E$
to be finite. It should instead diverge linearly with $r$. This
divergence is much milder, however, than in the case of flat spacetime
where the total average energy diverges as the volume of the system.

t\noindent{\bf \S 4. Discussion}

The spacetime outside of a sphere with temperature $T_s$ can thus be
divided into three regions. In the near field region, $r\ll r_m$ and
the solutions to Einstein's equations are given by $(19)$. The geometry
of the spacetime in this region is dominated by the mass $M$ at $r=0$
and the presence of the photons will not have an significant affect
on it. In the intermediate field region, $r_m< r < 1/\rho_o$ and the
total energy contained in the thermal photons is now comparable to
the mass of the sphere. Both the photons and the mass $M$ together will
determine the geometry of the spacetime. In the far field region,
$r\gg 1/\sqrt\rho_0$ and the asymptotic solutions $(26)$
are now valid. It is now the photons which are dominant over the
mass $M$.

We have in this paper considered only systems which are in
thermodynamic equilirbium. In particular, this means that the sphere
must be in thermodynamic equilibrium with the gas of photons
surrounding it. As the spacetime that we have been considering
consisted of only the mass $M$ and the photons, this is equivilant to
saying the sphere must be in thermodynamic equilibrium with the rest
of its universe. Since our universe is filled with with the cosmic
microwave background radiation which is at a temperature of $\sim
3^o$ K, for a physical body to be in thermodynamic equilibrium with
the rest of our universe it must also be at a compareable temperature.
There are very few actual physical bodies which are at such low
temperatures. Moreover, because the mass $M$ is in equilibrium with the
photons surrounding it, the amount of energy radiating away from the
sphere must exactly be balanced by the amount of energy impacting on the
sphere by the photon gas surrounding it. It is for this reason that
the energy density is dependent only on the geometry of the
spacetime, and is why the definition of the temperature $(12)$ makes
sense. It is also the reason why the intensity of the emitted
radiation from the sphere does not have the charactoristic $1/r^2$
behavior as one would naively expect.
\vfill
\supereject
\centerline{\bf Acknowledgements}

I would like to thank K.-W. Ng for many helpful discussions while
this paper was being written. This work is supported by the National
Science Council of the Republic of China under  contract number NSC
81-0208-M-001-78.
\vskip2truecm

\noindent{PACS numbers: 04.20.-q, 04.20.Jb, 04.40.+c, 5.90.+m
\vskip2truecm

\centerline{\bf REFERENCES}
\item{$[1]$}K. Schwarzchild, Kl. Math.-Phys. Tech., 189-196 (1916),
            Kl. Math.-Phys. Tech., 424-434 (1916).
\par
\item{$[2]$}N. D. Birrell and P. C. W. Davies, {\sl Quantum Fields in
                Curved Space}, Chapters 1,2 (Cambridge University
                Press, Cambridge 1982).
\par
\item{$[3]$}R. M. Wald, {\sl General Relativity}, Chapter 6, (The
                University of Chicago Press, Chicago, 1984).
\vfill
\end